\documentclass[final]{ias2}

\usepackage{graphicx} 
\usepackage{multirow}
\usepackage{array} 

\usepackage{hyperref} 
\usepackage{epsfig}
\begin{document}

\markboth{QCD Critical Point: The Race is On}
{Rajiv V. Gavai}

\title{QCD Critical Point: The Race is On
\vskip-2.5cm\hfill {\normalfont\large TIFR/TH/14-13}\vskip 2.2cm
}

\author[tifr]{Rajiv V.\ Gavai} 
\email{gavai@tifr.res.in}
\address[tifr]{Department of Theoretical Physics, Tata Institute of Fundamental
         Research,\\ Homi Bhabha Road, Mumbai 400005, India.}

\begin{abstract}

A critical point in the phase diagram of Quantum Chromodynamics (QCD), if
established either theoretically or experimentally, would be as profound a
discovery as the good-old gas-liquid critical point.  Unlike the latter,
however, first-principles based approaches are being employed to locate it
theoretically.  Due to the short lived nature of the concerned phases, novel
experimental techniques are needed to search for it.  The Relativistic Heavy
Ion Collider (RHIC) in USA has an experimental program to do so.  This short
review is an attempt to provide a glimpse of the race between the theorists and
the experimentalists as well as that of the synergy between them.

\end{abstract}



\maketitle

\section{Introduction}
\label{int}

Critical points in a phase diagram in the temperature-density plane are special
for several reasons. Starting from their rarity, only one or two in usual
systems, to the various dramatic changes in a host of physical quantities at
the critical point are the observations-based curiosities, while theoretically
the universality of critical indices, diverging correlation length etc. enable
us to relate widely different physical systems to the common underlying
symmetries.  For common substances, such as water or carbon dioxide, the
existence of critical point has been established experimentally, with its
location known rather precisely. Critical exponents for magnetic systems have
been found experimentally to be equal to those for the many liquid-gas systems,
and are related to the symmetry of the Ising model.  For many of the physical
systems, however, getting a theoretical, especially first principles based,
computation of their locations has turned out to be illusive still.  The aim of
this article is to describe how things are different for strongly interacting
matter, which is naturally described by Quantum Chromo Dynamics (QCD). QCD has
an inherently a higher energy, and thus a shorter time, scale.  Whether the QCD
phase diagram has a critical point in its temperature $T$ and the baryonic
chemical potential $\mu_B$ plane is therefore a tougher question to address
experimentally as well.  Thanks to the impressive developments on the
experimental fronts, the Relativistic Heavy Ion Collider (RHIC) at BNL, New
York, and the upcoming facilities such as FAIR at GSI, Darmstadt and NICA at
Dubna, Russia may be able to search for it, and potentially even locate it.

Unlike its cousin QED --quantum electrodynamics-- QCD turns out to be much
richer in structure.  While it can be handled theoretically in the (almost) same
way as QED when the coupling between its constituents quarks and gluon is
small, newer tools are needed in the strong coupling regime where a lot of the
rich structure, such as quark confinement, exists.  A variety of models have
been developed to handle this domain, and have been tested successfully for
hadronic interactions to a limited extent.  These models were was also the
first set of tools used for getting a glimpse of the QCD phase diagram
\cite{ErLy}.  For instance, using an effective chiral Nambu-Jana Lasinio type
model, a phase diagram was obtained, which suggests \cite{WilRaj} a critical
point to exist in a world with two light quarks (up and down flavours) and one
moderately heavier (strange) quark.  One would clearly like to have an {\em ab
initio} theoretical evidence for it from QCD directly.  This turns out to be
difficult as one has to deal with the strong coupling regime of QCD.
Non-perturbative Lattice QCD, defined on a discrete space-time lattice, has
proved itself to be the most reliable technique for extracting the information
from QCD  in the world of low energy hadronic interactions.  The hadron
spectrum has been computed and predictions of weak decay constants of heavy
mesons have been made \cite{FLAG13}.  Numerical simulations of lattice QCD
using the importance sampling approach have been the backbone for all such
computations. Since use of these techniques was actually inspired by their
success in statistical mechanics, it is natural to extend the same approach to
finite temperature and finite density QCD.  Indeed, its application to finite
temperature QCD has also yielded a slew of results for QCD thermodynamical
observables , such as the pressure as a function of temperature.  It is
therefore natural to ask whether lattice QCD can help us in locating the QCD
critical point.  The prospect of finding the critical point experimentally
makes it exciting both as a check of such theoretical predictions and as a
competition for getting there first.  Clearly, in view of the complexity of the
task, one could turn to lattice QCD again to see if it can provide any hints
for the experimental search program as well.  In this short review, I aim to
summarise the results obtained in this direction, including those by the TIFR
group, Mumbai. 

Lattice QCD simulations at finite density, or equivalently nonzero $\mu_B$,
have had to face to two severe problems, which have resulted in a slow progress
in the field of QCD critical point.  Recently, one of them, a conceptual issue,
appears to have been solved.   Let me explain that first.  Due to the
well-known fermion doubling problem, one has to make a compromise in choosing
the quark type for any lattice QCD computation.  While the final results should
be independent of the quark type, indeed any form of the lattice action, the
choice of the quark type is often dictated by the nature of the task at hand.
The popular choice in finite temperature and density studies has been mostly
the Kogut-Susskind (staggered) quarks \cite{LaTex}.  They have an exact chiral
symmetry at any value of the lattice spacing (any finite cut-off) which leads
to an order parameter, the chiral condensate $\langle \bar \psi \psi \rangle$,
for a lattice investigation in the entire $T$-$\mu_B$ plane. Unfortunately
though, the flavour and spin symmetry is broken for them on a lattice. A
simulation relevant to experiments ought to have  2 light (or 2 light + 1
heavier ) quark flavours.  Precisely such a representation for staggered quarks
may not be feasible, more so on the coarse lattices one is constrained to
employ.  The existence of the critical point, on the other hand, is expected to
depend crucially on the number of flavours.  Flavour and spin symmetry is
therefore needed in simulations.  Although computationally much more expensive,
domain wall or overlap fermions \cite{DwOv} are better in this regard, as they
do have the correct symmetries for any lattice spacing at zero temperature and
density.  Introduction of chemical potential, $\mu$, for these is, however, not
straight-forward due to their non-locality.  Bloch and Wettig \cite{BlWe}
proposed a way to do this.  Unfortunately, it turns out \cite{BGS} that their
prescription breaks chiral symmetry.  Of course, a space-time lattice breaks
even translational and rotational symmetry.  Exactly as these are expected to
be restored in the continuum limit of vanishing lattice spacing, one may argue
that the same would be the case for the Bloch-Wettig method.  The problem is
that for the chiral condensate to be an order parameter on a lattice, the
chiral symmetry must be respected even on the lattice.   Otherwise cut-off
dependent corrections to it would change it in an uncontrollable way.
Furthermore, the chiral anomaly for the Bloch-Wettig action depends on $\mu$
unlike in continuum QCD \cite{GS10}. As a result, it too may alter the
existence/location of the critical point on the lattice. Recently, this
conceptual issue has been resolved \cite{GS12} by a new proposal to introduce
the chemical potential for both overlap and domain wall fermions.  It leads to
a formalism with both a) continuum-like (chiral, flavour and spin) symmetries
for quarks at nonzero $\mu$ and $T$ and b) a well-defined order parameter on
the lattice.  

Finite density simulations needed for locating a critical point suffer from
another well known problem.  It is inherited from the continuum theory itself:
the fermion sign problem.  It appears in areas of statistical mechanics also and
has so far remained unsolved.  It is a major stumbling block in extending
the lattice techniques to the entire $T$-$\mu_B$ plane.  Simply stated,
simulations rely on importance sampling due to the large number of field (quark
and gluon) variables.  Assuming $N_f$ flavours of quarks, and denoting by
$\mu_f$ the corresponding chemical potentials, the QCD partition function is 

\begin{equation}
{\cal Z} =  \int \scriptscriptstyle{D} U \exp(-S_G) ~ \prod_{f} {\rm
Det}~M(m_f, \mu_f)~~, 
\end{equation}   
and the thermal expectation value of any observable ${\cal O}$ is
\begin{equation}
 \langle {\cal O} \rangle = \frac{   \int \scriptscriptstyle{D}
U \exp(-S_G) ~ {\cal O} \prod_{f} {\rm Det}~M(m_f, \mu_f)} { {\cal Z~~}} .  
\end{equation}  

For a $N_s^3 \times N_t$ lattice, the simulations then correspond to a spatial
volume of $V= N_s^3a^3$, and a temperature of $T = (N_ta)^{-1}$, where $a$ is
the lattice spacing.  Quark masses and chemical potentials are then also in
units of $a$.  Typically, the integral above is ~ 10 million dimensional and
$M$, the fermion matrix defined by the lattice Dirac operator, is about a
million $\times$ million.  Probabilistic methods are therefore used to evaluate
$\langle {\cal O} \rangle$. These need  a positive definite measure.  From
Eq.(1), one sees the measure to be proportional to the exponential of the
gluonic action, $\exp(-S_G)$ and the fermionic determinant Det$~M$.  While the
former factor is positive definite, for nonzero $\mu$, the determinant turns out
to be a complex number, although the partition function itself remains real.
This is the fermion phase/sign problem in finite density QCD.  Note that the
determinant is purely real for $\mu=0$.  Several approaches have been proposed
in the past to deal with it.  In view of the space constraints, let me provide a
partial list, with some comments:  

\begin{enumerate} 

\item Suitable choice of variables \cite{Cha13,Gat14}--- Since the partition is
still real, one can attempt suitable variable transformations to re-write it
using variables which do not lead to a complex measure.  The key here is, of
course, to find such variables.  This approach has been shown to work in simpler
models or with some drastic approximations to QCD but so far remains a distant
goal for realistic QCD simulations.

\item Imaginary chemical potential \cite{ImMu} --- Substituting $\mu \to i \mu$
in the fermion determinant, one can easily verify that it becomes real.  Usual
simulation methods then enable one to explore the imaginary $\mu$-axis.
However, the results so obtained can only be employed for our world after
analytic continuation, which introduces unknown and uncontrollable errors.  So
far, it has been used for very coarse ($N_t=4$) and with only the leading term
in an polynomial ansatz for analytic continuation.  No hints of QCD critical
point were seen.  It will be interesting to see how these results will be in the
continuum limit and with more terms in the polynomial.

\item Canonical ensemble method \cite{CaEn} --- Computations at a fixed baryon
number do not have the fermion sign problem as well. Using this idea,
simulations on small lattices have been done with encouraging results.  One
needs to compute them in the limit of very large baryon number in order to
obtain the corresponding chemical potential. Until that is achieved, one can
only make qualitative comparisons.  The hints of a first order phase transition
seen in these simulations are consistent with a possible critical point at some
other, perhaps lower, $\mu$.

\item Complex Langevin approach \cite{ComLa} --- Stochastic quantization of
field theories aims to evaluate the  $\langle {\cal O} \rangle$ by formulating
the problem in terms of a Langevin equation and by employing its (simulation)
time evolution.  It too, of course, needs the same reality and positivity
condition.  Although, a formal proof does not exist for complex action, one may
still hope that it converges to the correct answer.  Such attempts for some
models have been shown to work, i.e., they agree with the known
exact/semi-analytic results but in some cases it is known to fail as well.
Currently attempts are being to made to understand this to see if it can be
eventually applied to full QCD.

\item Two parameter re-weighting \cite{FoKa}  --- Importance sampling, basic to
many of the simulation methods, works by creating the most probable distribution
of the action in Eq. (1) for given input parameters, such as the gauge coupling.
Exploiting cleverly the fact that such a distribution can be constructed for
$\mu=0$ by using standard methods, this approach consists of obtaining the
desired $\mu \ne 0$ distribution from the expectation values with respect to the
former.  Fodor and Katz \cite{FoKa} were the first to obtain a lattice result
for the QCD critical point on small and coarse lattices ($4^3 -8^3 \times 4$),
which was found to be at $\mu_E/m_{Nucleon}\sim 0.77$.  Employing similar
lattice but a realistic quark mass spectrum in \cite{FoKa}, they obtained a
critical point at a factor 2 lesser $\mu_E$.  Extending this approach to larger
and fine lattices appears not to be easy because of an exponential growth in
the size of the computational resources needed due to the so-called `overlap'
problem but would be nice if it could be done.

\item Taylor expansion \cite{TaExM, TaExB} --- The Taylor expansion approach was
developed by the TIFR group \cite{TaExM} and the Bielefeld group
\cite{TaExB}. Results obtained by this method have been rather encouraging, as
one has been able to study systematically i) the finite volume effects to go
towards a thermodynamical limit ($N_t \to \infty$) and ii) the continuum limit
of vanishing lattice spacing ($N_t \to \infty$ or equivalently $a \to 0$).  I
will discuss some details of these results in the next section.

\end{enumerate}

\section{Lattice Results}
\label{olr}

Expanding the partition function of Eq. (1)  near $\mu_f=0$, where $f$ is the
flavour index denoting up, down, strange..quarks, one can evaluate the
coefficients of the various terms by employing the usual simulation method since
they are evaluated for $\mu_f=0$ for which the determinant is real.  Moreover,
both the thermodynamic limit, $N_s \to \infty$, and the continuum limit, $N_t
\to \infty$ of these coefficients can be evaluated in the same way as the hadron
spectrum can be computed in these limits to obtain the continuum results.  We
will consider below only two flavours, up and down, with the corresponding
chemical potentials, $\mu_u$ and $\mu_d$ for simplicity. Generalizing to any
number of flavours is straightforward, although for our world two light and one
moderately heavy flavour may suffice. 

Using standard definitions, number densities $n_i$ and the corresponding
susceptibilities $\chi_{ij}$ can be obtained from Eq. (1) as respectively the
first and second derivatives of the partition function ${\cal Z}$ :

\begin{equation}
n_i =\frac{T}{V}\frac{\partial \ln {\cal Z}}{\partial \mu_i}|_{T=\text{fixed}}
~;
\qquad
\chi_{ij} = \frac{T}{V} \frac{\partial^2 \ln {\cal Z}}{\partial\mu_i \partial \mu_j}|_{T=\text{fixed}} ~. 
\end{equation} 

Natural units with the Boltzmann constant $k$ set to unity have been used above.
While the $n_i$ are zero for $\mu_i=0$, the susceptibilities are a nontrivial
function of temperature.  Indeed, these are known to play many roles in the
physics of finite temperature phase transition, including acting as independent
indicators of the transition itself.  Denoting by $\chi_{n_u,n_d}$ higher
derivatives of $\ln {\cal Z}$ at the $n_u^{\rm th}$ ( $n_d^{\rm th}$) order with
respect to $\mu_u$ ($\mu_d$), the QCD pressure $P =  T \ln {\cal Z}/V $
is seen to have the following expansion in $\mu_i$ :
\begin{equation}
   \frac{\Delta P}{T^4} \equiv \frac{P(\mu, T)}{T^4} - \frac{P(0, T)}{T^4}
   = \sum_{n_u,n_d} \chi_{n_u,n_d}\;
        \frac{1}{n_u!}\, \left( \frac{\mu_u}{T} \right)^{n_u}\, 
        \frac{1}{n_d!}\, \left( \frac{\mu_d}{T} \right)^{n_d}\, 
\end{equation} 

At a generic point $(\mu, T)$ in the QCD phase diagram, one should be able to to
sum the series for the pressure as a function of temperature $T$.  In presence
of a critical point at $T_E$, however, a finite radius of convergence in $\mu$
would limit the range up to which such a summation can be done.  A typical
lattice computation can thus aim to search for the QCD critical point by looking
for the radius of convergence of the series at several temperatures.   Since the
baryonic susceptibility should have a stronger power law divergence at the
critical point, it was proposed \cite{our1} to use that series instead of that
of pressure above.  It is easy to construct the series for baryonic
susceptibility from the above expansion and look for its radius of convergence
as the estimate of the nearest critical point.   

Due to the symmetry between baryons and antibaryons at $\mu = 0$, odd
coefficients of the series vanish.  Using the even ones, successive estimates
for the radius of convergence can be obtained by using the usual ratio method
\begin{equation}
r_{n+1,n+3} = \sqrt{\frac {n(n+1)\chi^{(n+1)}_B} {\chi^{(n+3)}_B T^2}}  
\label{nrati}
\end{equation}
and the
n$^{th}$ root method 
\begin{equation}
 r_n = \Bigg(\frac{n!{\chi_B^{(2)}}}{{\chi_B^{(n+2)} T^n}} \Bigg)^{1/n}~. 
\label{nthrt}
\end{equation}
Of course, one needs to evaluate the $n \to \infty$ limit of the radii
estimates above to arrive at the true radius of convergence.  Due to the
computational complexity involved, both the number of terms and cancellations
amongst these terms increase as $n$ grows, terms up to 8th order in $\mu$ have
been used so far \cite{our1}.  A proposal \cite{GS10}  to alleviate these
problems, and thus get more coefficients for the same computational resources
has been made. It consists of adding the chemical potential simply as $\mu
N$\cite{GS10}, in contrast the commonly used form exponential in $\mu$.   The
linearity in $\mu$ implies that most derivatives of the quark matrix with $\mu$
are zero except the first one.  This reduces a lot the number of terms in each
nonlinear susceptibility (NLS) appearing in the equations above.  One can show
that it still leads to essentially the same results\cite{gs1}.  It would be
interesting to see the results for the radius of convergence obtained by using
it.  

Since the QCD critical point is expected to occur for real $\mu$, a
key point to note is that all the coefficients of the series must be positive.
Having made sure that it is so, one can then look for agreement between the two
definitions above as well as their $n$-independence to locate the critical
point. The detailed expressions for all the terms can be found in \cite{our1}
where the stochastic estimators method to evaluate them is also explained.  For
terms up to the 8$^{th}$ order one needs 20 inversions of the Dirac matrix, $(D
+m)$, on $\sim$ 2000 vectors for a single measurement on a given gauge
configuration.  At least hundreds of such measurements are necessary to obtain
the results discussed below; more precise results will need even more.
This makes the computation very time consuming.  Nevertheless, extension
to 10th \& even 12th order may be possible, using the idea in \cite{GS10} which
may save up to 60 \% computer time in these measurements. 

The results of Ref. \cite{our1,our2,our4} were obtained by simulating full QCD
with two flavours of staggered fermions of mass $m/T_c =0.1$ on $N_t \times
N_s^3$ lattices, with $N_t=4$ and $N_s=$ 8, 10, 12, 16, 24 \cite{our1}, a finer
$N_t= 6 $ with $N_s=$ 12, 18, 24 \cite{our2} and a still finer $N_t=8$ and $N_s
=32$ lattice. Here $T_c$ denotes the transition temperature at $\mu=0$.  It is
obtained by studying the point of inflection in variables like the chiral order
parameter or the Polyakov line.  Earlier work by the MILC collaboration for the
smaller $N_t = 4$ lattice had determined \cite{milc} the ratio of the mass of
the rho particle to the transition temperature, $m_\rho/T_c = 5.4 \pm 0.2$ and
the similar ratio for pion, $m_\pi/m_\rho = 0.31 \pm 0.01$. This amounts to a
Goldstone pion of 230 MeV in that case, which is heavier than the physical pion
mass of 140 MeV.  For the finer lattices, Ref. \cite{our2,our4} had to determine
the $\mu=0$ transition points $\beta_c$ as well.  It was checked that these were
consistent with the nearby quark mass results in the literature.  Covering
typically a temperature range of 0.89 $ \le T/T_c \le $  1.92  by suitably
choosing the range of couplings on all lattices, the coefficients were
determined  typically on 50-200 independent configurations, separated by the
respective autocorrelation lengths.

\begin{figure}[htb]
\includegraphics[scale=0.48]{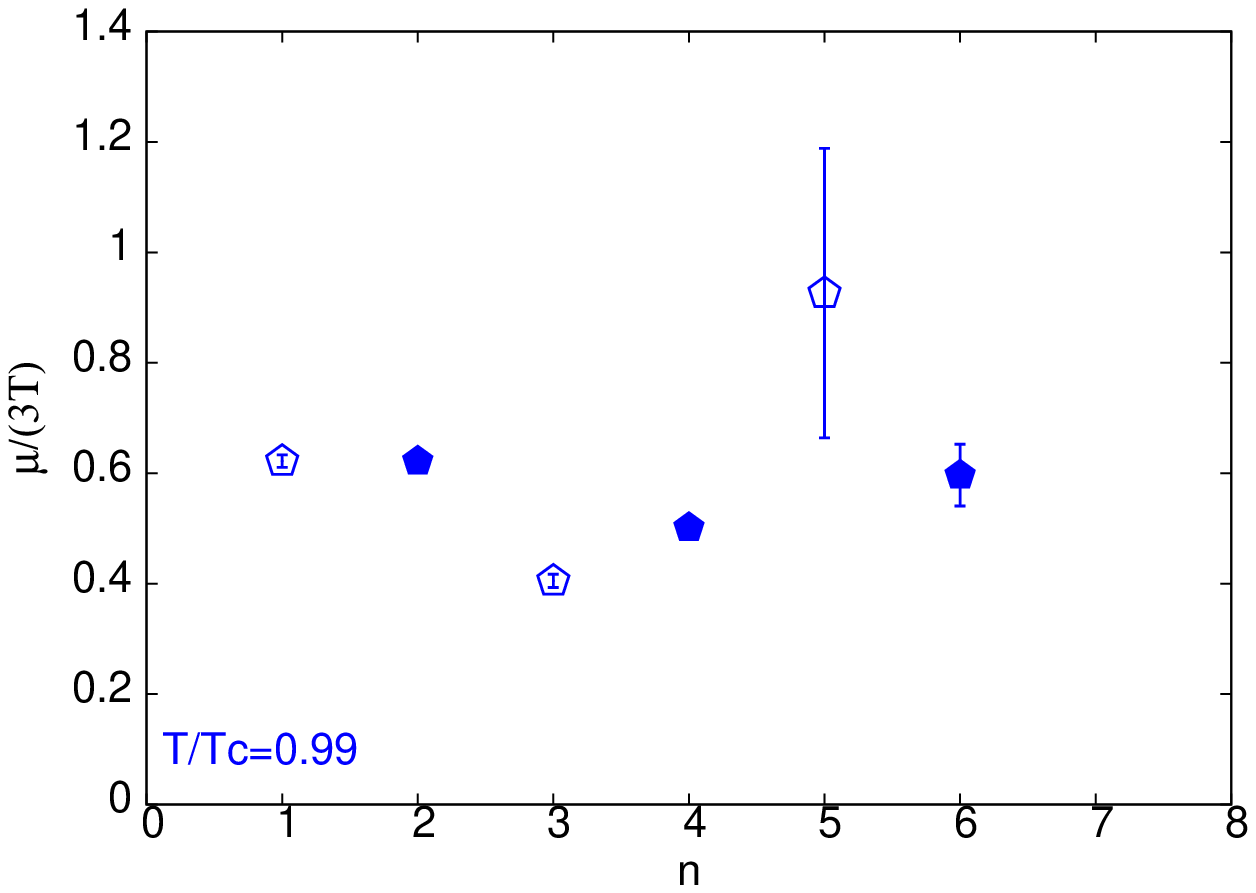}
\includegraphics[scale=0.48]{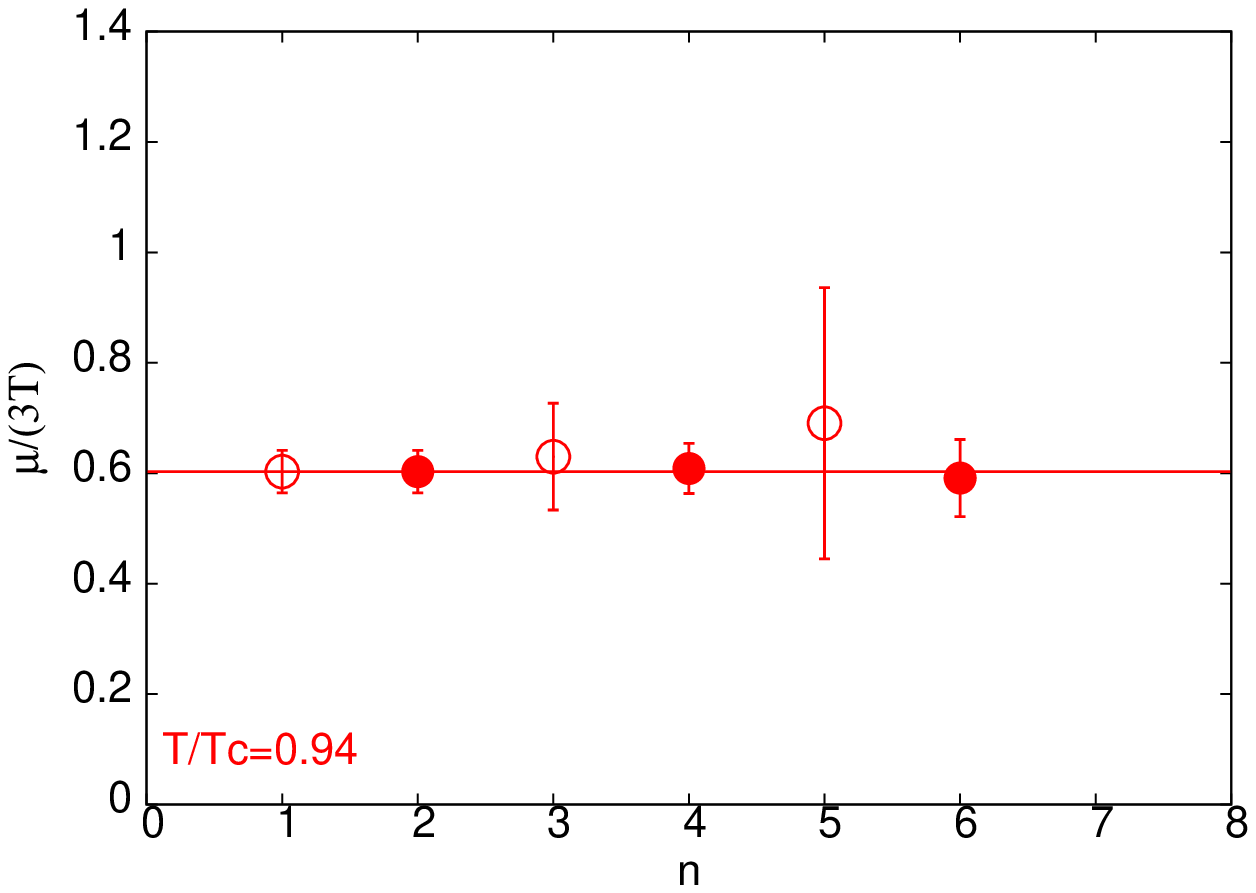}
\caption{Estimates of radii of convergence as a function of the order $n$ and
for the two methods (see text) at $T/Tc$ = 0.99 and the critical point $T^E/T_c$
=0.94.  The open (filled) points display the results for the ratio (root)
method.  From Ref. \cite{our2}.  }
\label{rads}
\end{figure}

Fig. \ref{rads} shows the results \cite{our2} for both the ratios defined above
on the $N_t=6$ lattice at two different temperatures, $T/T_c =0.99$ and 0.94.
In order to exhibit both the results of eqs. (\ref{nrati}-\ref{nthrt}) the $n$
on the x-axis was suitably transformed in those equations such that the hollow
(filled) symbols show the ratios($n^{\rm th}$ root).  All the susceptibilities
are positive at both the temperatures but the ratios fluctuate for the former
temperature while seeming to be constant for the latter, indicating it to be
the critical temperature and their constant value to be the $\mu_u/T = \mu_d/T
= \mu_B/3T$.  Ref.  \cite{our2} thus found the coordinates of the endpoint
(E)---the critical point---to be $ T^E/T_c = 0.94\pm0.01$, and $\mu_B^E/T^E =
1.8\pm0.1$.

\begin{figure}[htb]
\includegraphics[scale=0.48]{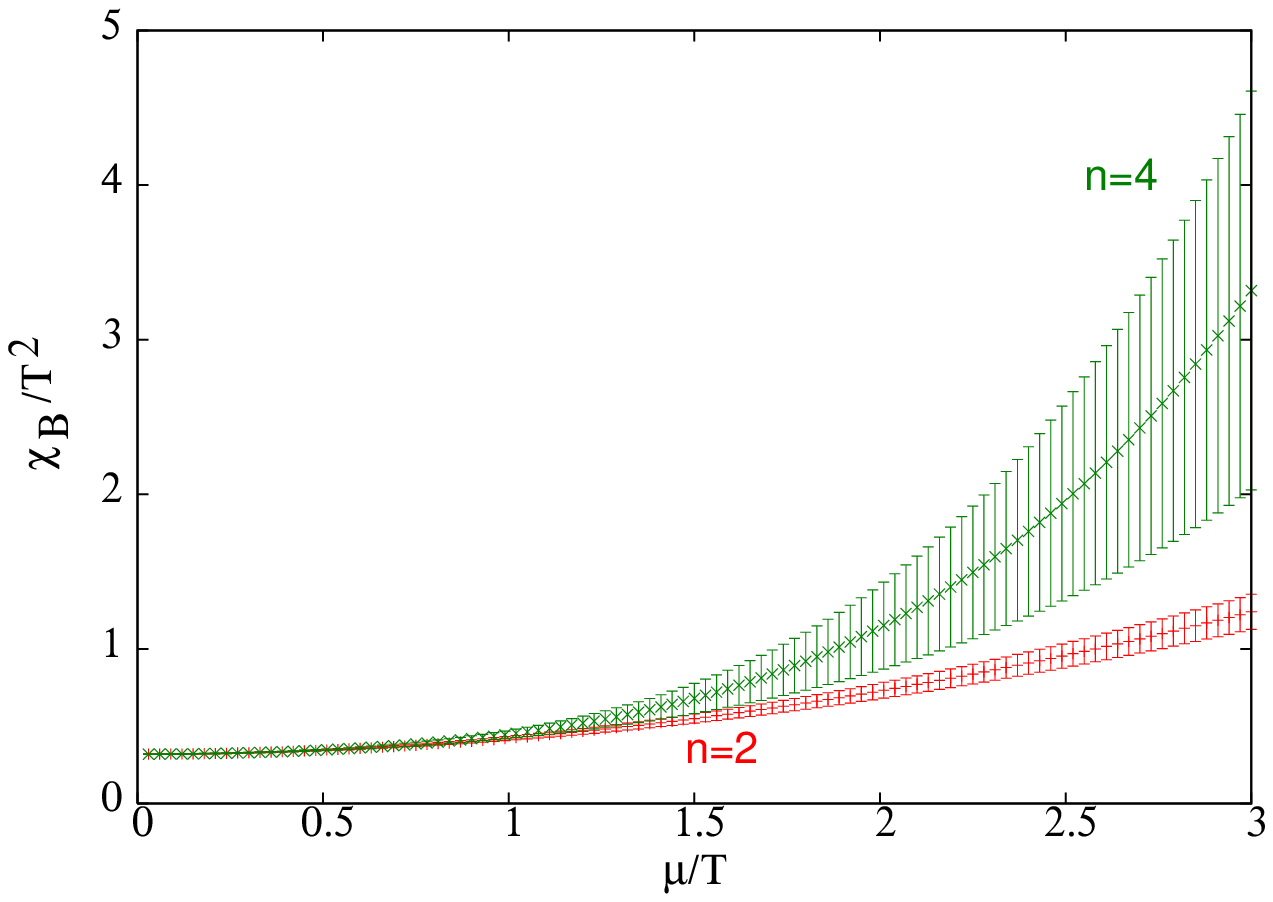}
\includegraphics[scale=0.48]{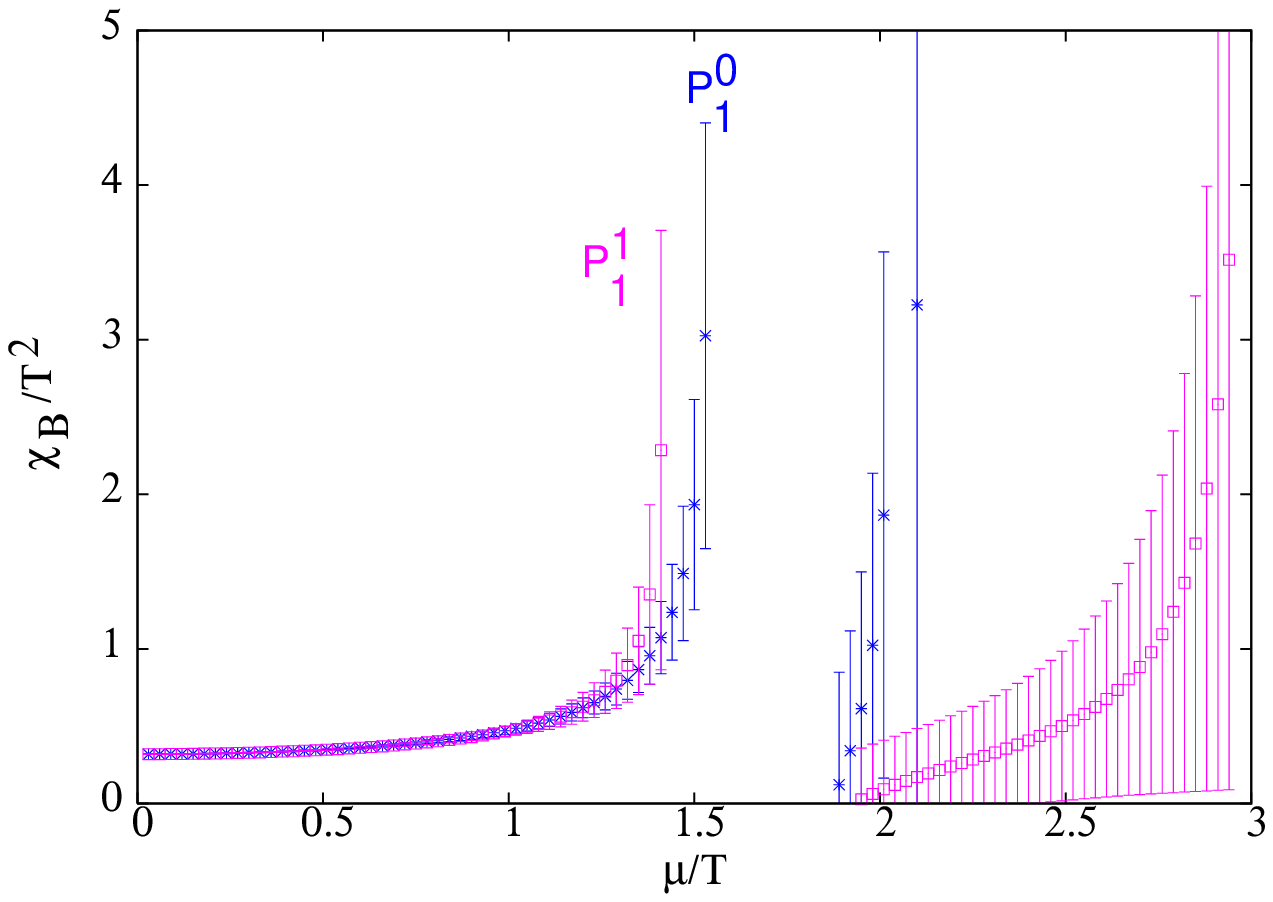}
\caption{The baryonic susceptibility as a function of the baryonic chemical
potential using simple polynomial with all computed orders (left) and Pad\'e
approximants (right).  From Ref. \cite{our2}. }
\label{polypad}
\end{figure}

As a cross check on the location of the $\mu^E/T^E$, one can use the  series
with the same numerically determined coefficients to construct $\chi_B$ for
nonzero $\mu$ directly.  As one sees in the left panel of Fig. \ref{polypad},
this leads to smooth curves with no signs of criticality irrespective of how
many terms one uses to determine the curve.  This is, of course, to be expected,
since the series is merely a finite (and small) order polynomial.  It is well
known from statistical mechanics that Pad\'e approximants are a better guide in
such cases. Employing the Pad\'e approximants for the same series to estimate
the radius of convergence does lead to a consistent window with the above
estimate, as seen the right panel of  Fig. \ref{polypad}. 

\begin{figure}[htb]
\includegraphics[scale=0.48]{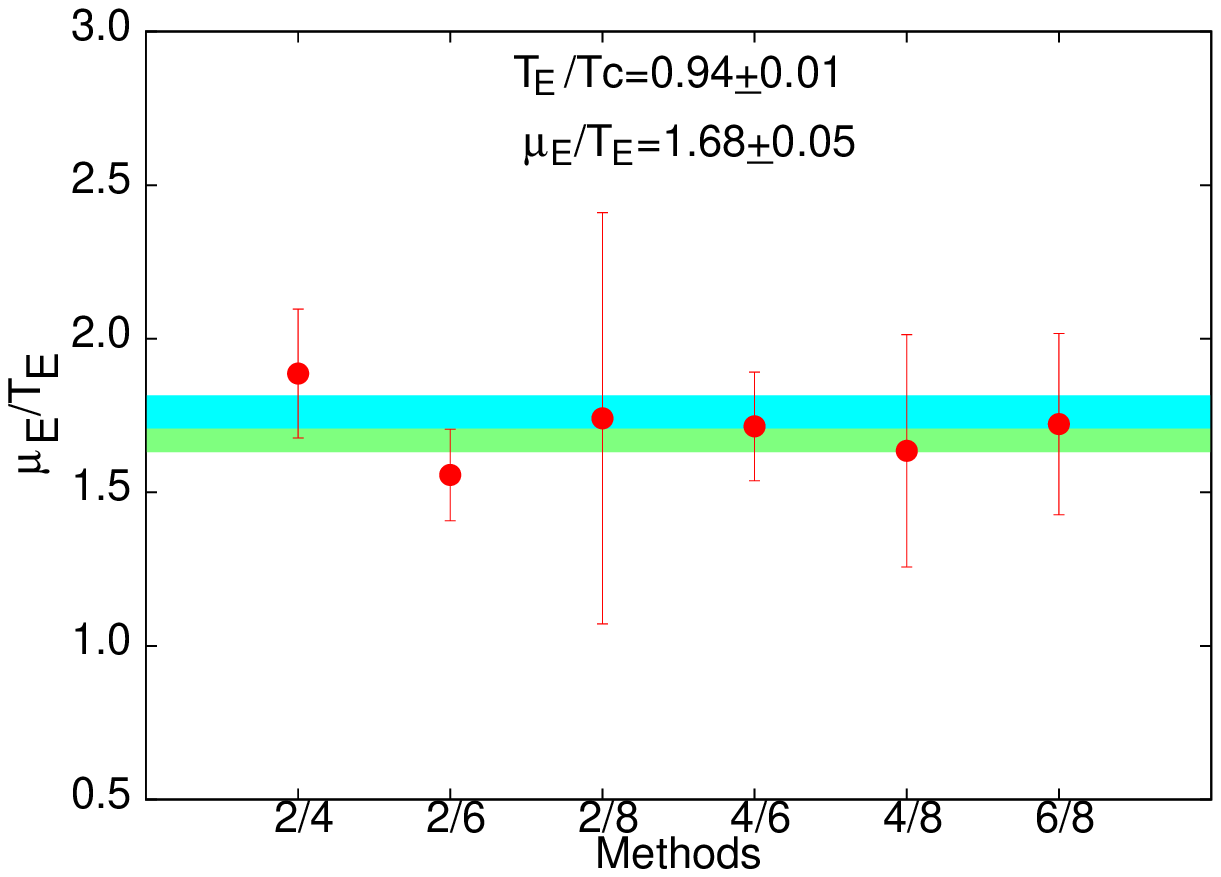}
\includegraphics[scale=0.48]{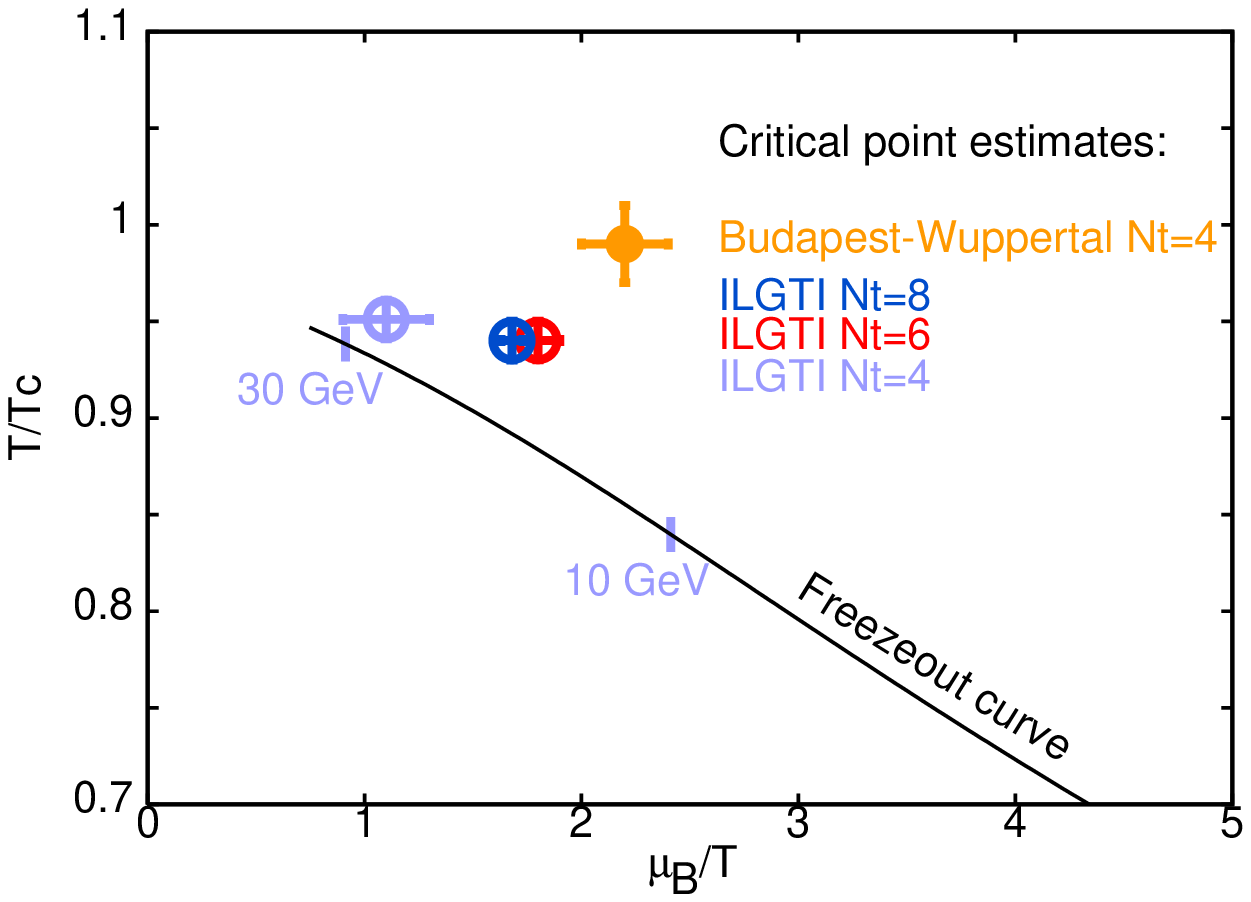}
\caption{Estimates of radii of convergence as a function of the ratios used for
the two methods at the critical point $T^E/T_c$ for $N_t =8$ (data) along with
the final results \cite{our4} for $N_t =$ 8 (green band) and 6 (cyan band)
lattices.  The right panel shows the QCD phase diagram from known lattice
determinations \cite {FoKa,our1,our2,our4}. See text for details. }
\label{CrPt}
\end{figure}

The left panel of the Fig. \ref{CrPt} displays a comparison of the results on
the even finer $N_t =8$ lattice \cite{our4} with those for the $N_t =6$ lattice.
Significantly more computation power is needed for the larger lattice not only
because the computations grow as $N_s^3 \times N_t$ but also cancellations of
various terms become more delicate.  As in the cases of the smaller $N_t =4$ and
6 lattices, the data for the appropriate radii-estimates for $\mu_E/T_E$ exhibit
near constancy for the various possible ratios of the Taylor coefficients.  The
first (last) three data point display the $n^{\rm th}$ root (ratio) results.
The two bands are the final estimates for $N_t =6$ (cyan) and 8 (green) lattices
\cite{our2,our4}.  One sees that they agree within errors whereas the earlier
coarsest \cite{our1} lattice result was at a somewhat smaller $\mu_B/T$.  The
right panel of Fig. \ref{CrPt} shows the QCD phase diagram with known lattice
determinations \cite {FoKa,our1,our2,our4}.  Given the complexity of the problem
it is very encouraging that a) two different methods with their own different
shortcomings and 2) three different $N_t$ or equivalently three different
cut-offs lead to nearby determinations of the location of the critical point.
The agreement within error bands for the two finest lattices is further
encouraging.  One has to bear in mind though that there is still quite a long
way to go.  Cut-off independence (still larger $N_t$), volume independence
(larger aspect ratio $N_s/N_t$ than 4 employed here) as well as the independence
from other choices, such as the type of quarks or method have to be further
established.  The simulations also need to come closer to the continuum QCD by
having physical pion mass, i.e., lower dynamical quark mass, and the correct
$U_A(1)$ anomaly.  Nevertheless, one can hope that the concurrence of these
results can already motivate strong efforts to locate the critical point
experimentally, and as we shall see below, even provide some guidance on how to
decipher that from the hadronic data visible to the detectors.

\section{Searching Experimentally}
\label{se}

Searching for the QCD critical point experimentally also entails several
formidable challenges, in view of the anticipated high values of its temperature
and density. Bjorken \cite{Bj82} argued that matter at such temperatures may be
created in relativistic heavy ion collisions. Indeed, major experimental efforts
at BNL, New York and CERN, Geneva in the past few decades have yielded
spectacular results \cite{HvRvs} supporting such an idea.  It was observed that
huge amounts of precise data on hadron abundances in a range of relativistic
heavy ion experiments at several colliding energies ($\sqrt{s}$) are well
described \cite{stm}  by statistical thermodynamical models with minimal
parameters such as temperature $T$ and the baryonic chemical potential $\mu_B$.
It led to a mapping of the $T$ and $\mu_B$ parameters of the model to the
collision (center of mass) energy $\sqrt{s}$, the so-called freeze-out curve.
One such parameterization of the freeze-out curve is shown in the QCD phase
diagram in the right panel of Fig. \ref{CrPt}.   Although, this curve can, on
one hand, be deemed as a mere parameterization of the hadron yields in the
heavy ion collision experiments, it has the advantage of being solely obtained
from precise experimental data and provides the $(T, \mu)$ accessible in these
heavy-ion experiments as a function of the colliding energy $\sqrt{s}$.
Identifying them as the thermodynamic variables of the fireball created
therefore gives us an idea of the likely colliding energy one may need to
employ to explore the region where the QCD critical point may be expected.  One
expects it to be at rather small colliding energies than the original RHIC
design of 200 GeV or the LHC energy.  The encouraging aspect of the lattice
results in Fig. \ref{CrPt} is that the expected range of $\sqrt{s}$ is not too
low, as would be the case if the critical point were to be at too high a
$\mu_B$.   It turned out that the RHIC accelerator could be tuned to run at
lower $\sqrt{s}$, a technologically impressive feat since it was not designed
for that.  Considering the spread in the values of lattice $\mu_E/T$ in Fig.
\ref{CrPt}, a beam energy scan program was proposed and sanctioned.   I shall
mostly concentrate below on some of the results of these program.  For more
details, I refer the reader to the experimental review in this volume.

Since the well-known  experiments of Thomas Andrews in 1869 for the liquid-gas
transition in carbon dioxide, one expects critical opalescence, or huge
fluctuations, to characterise the second order phase transition at the critical
point.  A diverging correlation length $\xi$ is now understood to be the cause
for this phenomenon, and is related to the diverging specific heat or
susceptibility at the critical point.  Stephanov, Rajagopal and Shuryak
\cite{srs99} proposed that studying the event-by-event fluctuations of suitably
chosen observables in heavy ion collisions as a function of $\sqrt{s}$) and
suitable kinematic cuts can be used to search for the QCD critical point due to
the nonmonotonicity a critical point should induce in them. They proposed to
focus on the fluctuations, or the second moments, of the observables constructed
from the multiplicity and transverse momenta of charged pions.  These have been
investigated with no conclusive results.  Recognising that the higher moments of
the same variables grow as a significantly higher power of $\xi$, Stephanov
\cite{ms09} suggested fourth and higher moments to be a more attractive tool. 

Considering that the QCD critical point arises due to a variation in the baryon
density, such fluctuations in baryon number, given by the appropriate
combinations of the many quark number susceptibilities discussed in the previous
section, appear a more natural choice \cite{sg09,our3}.  In order to make
predictions for the heavy ion experiments, and design further any search
criteria for the critical point, it was proposed \cite{our3} to use the
freeze-out curve as a a way to set the temperature $T$ and the chemical
potential $\mu_B$ in the lattice QCD computations to compute the various
fluctuations.  The information hidden in the nonlinear susceptibilities
discussed in the previous section can then be extracted by evaluating lattice
QCD predictions along it.  Not only does such an approach amounts to assuming
that the  $(T, \mu)$ along the freeze-out curve do reflect the true
thermodynamic variables of the system, but it also takes the model \cite{stm} a
step further and aims to predict the fluctuations in the hadron abundances from
first principles.   Comparing these lattice predictions with data even at large
$\sqrt{s}$, away from the critical point, thus is tantamount to a
non-perturbative test of QCD, subject to the assumptions outlined already.  

One expects the QCD critical point to have a critical region whose size depends
on the size of the fireball as well as its critical exponents.  The freeze-out
curve may pass through it or may miss it.  In the former case, one should expect
the fluctuations to be enhanced in the experiments but nothing spectacular may
happen in the latter case.  If the conditions are right, it may even pass close
enough to the critical point such that a study of fluctuations along it will
detect its presence unambiguously.   Defining $ m_1 =
T\chi^{(3)}(T,\mu_B)/\chi^{(2)}(T,\mu_B)$, $ m_3 =
{T\chi^{(4)}(T,\mu_B)}/{\chi^{(3)}(T,\mu_B)}$, and $ m_2=m_1m_3$, and noting
that the variance, skewness and kurtosis of the event distribution of baryon
number measure the various $\chi$'s appearing in them, one sees that these
ratios can be computed from lattice studies as well as from the experimental
data.  They are designed \cite{sg09, our3} such that the spatial volume cancels
out in them, making them suitable for both i) lattice studies, which can
potentially have finite volume effects, and ii) experiments which too are unable
to pin down the volume of the fireball and may prefer to use their favourite
proxy for it.  Usually, number of participants is preferred for that role in the
analysis of heavy ion data.  All lattice results also have to worry about the
finite lattice cut-off effect; one has to take the cut-off away (lattice spacing
$a \to 0$ or $N_t \to \infty$) by some extrapolation to obtain continuum
results.  Since dimensionless ratios have generally found to vary rather slowly
with $a$ for small enough $a$, the dimensionless $m_i$'s are also expected to be
reasonably cut-off independent.

Defining $z = \mu_B/T$, and denoting by $r_{ij}$ the estimate for radius
of convergence using $\chi_i$, $\chi_j$ as before, one has 

$$m_1 = \frac{2z}{r_{24}^2} \big[ 1 + \big(\frac{2r_{24}^2}{r_{46}^2} -1 \big) 
z^2 + \big(\frac{3r_{24}^2}{r_{46}^2 r_{68}^2} - \frac{3r_{24}^2}{r_{46}^2} +1
\big) z^4 + {\cal O}(z^6) \big]~.~$$
Note that the coefficients of the polynomial above are determined by the zero
density ($\mu =0$) simulations as described in the previous section and depend 
only on temperature.  Similar series expressions can be written 
down for $m_2$ and $m_3$ as well. In order to capture the critical behaviour,
where at some $T_E$ all these radii tend to be equal, a simple series summation
may be inadequate, as in Fig. \ref{polypad}.  One should resum these by the 
Pad\`e method as above to construct $m_i(\sqrt{s}) \equiv m_i(z=\mu_B/T, T)$, 
since ) the Pad\`e approximants seem to capture the critical behaviour well 
in Fig. \ref{polypad}, and ii) computations along the freeze-out curve
guarantee the equivalence for $m_i$.  A suitable ansatz is : 

$$ m_1 = z P^1_1(z^2;a,b), \qquad m_3 = \frac{1}{z} P^1_1(z^2;a',b')$$.

\begin{figure}[htb]
\begin{center}
\includegraphics[scale=0.48]{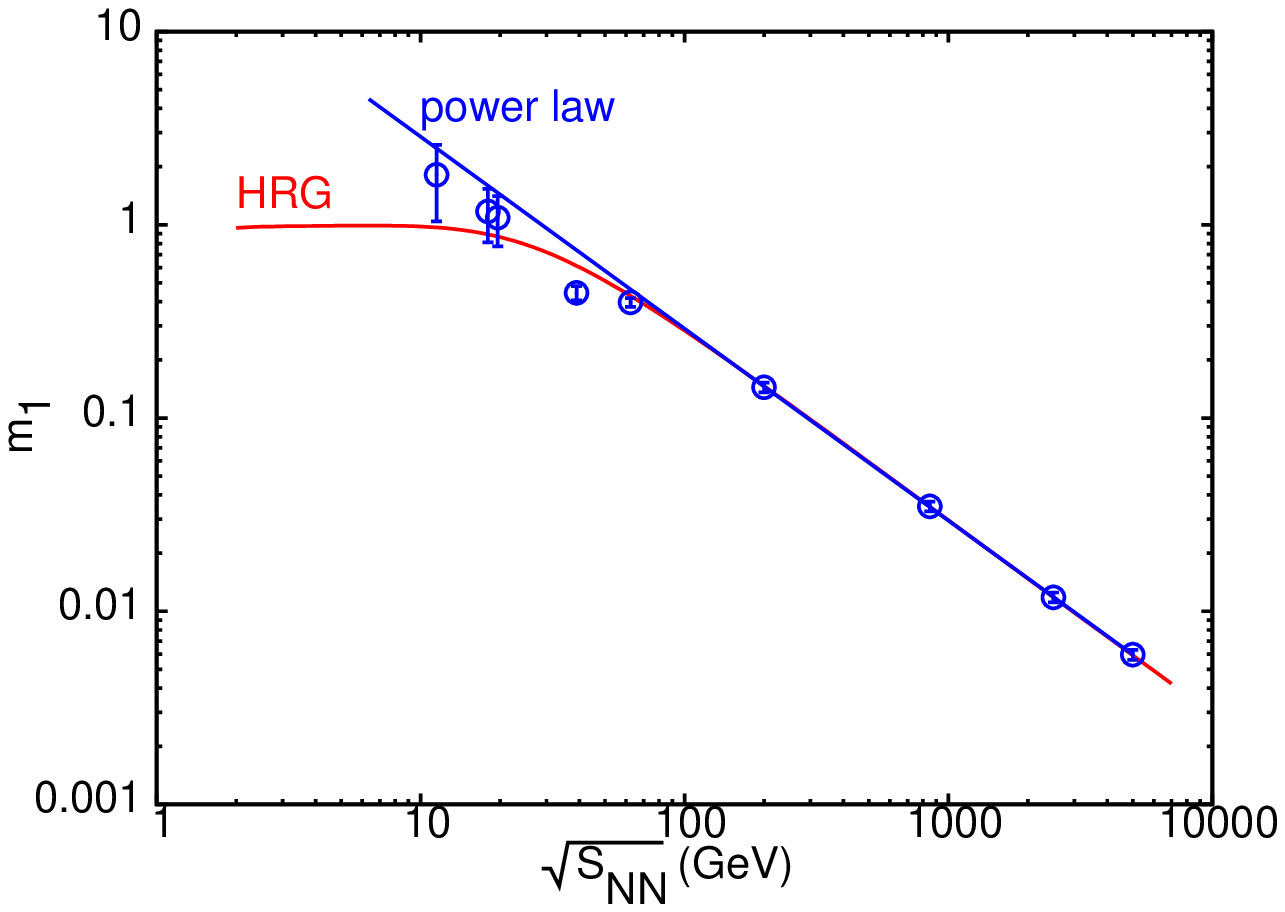} 
\includegraphics[scale=0.48]{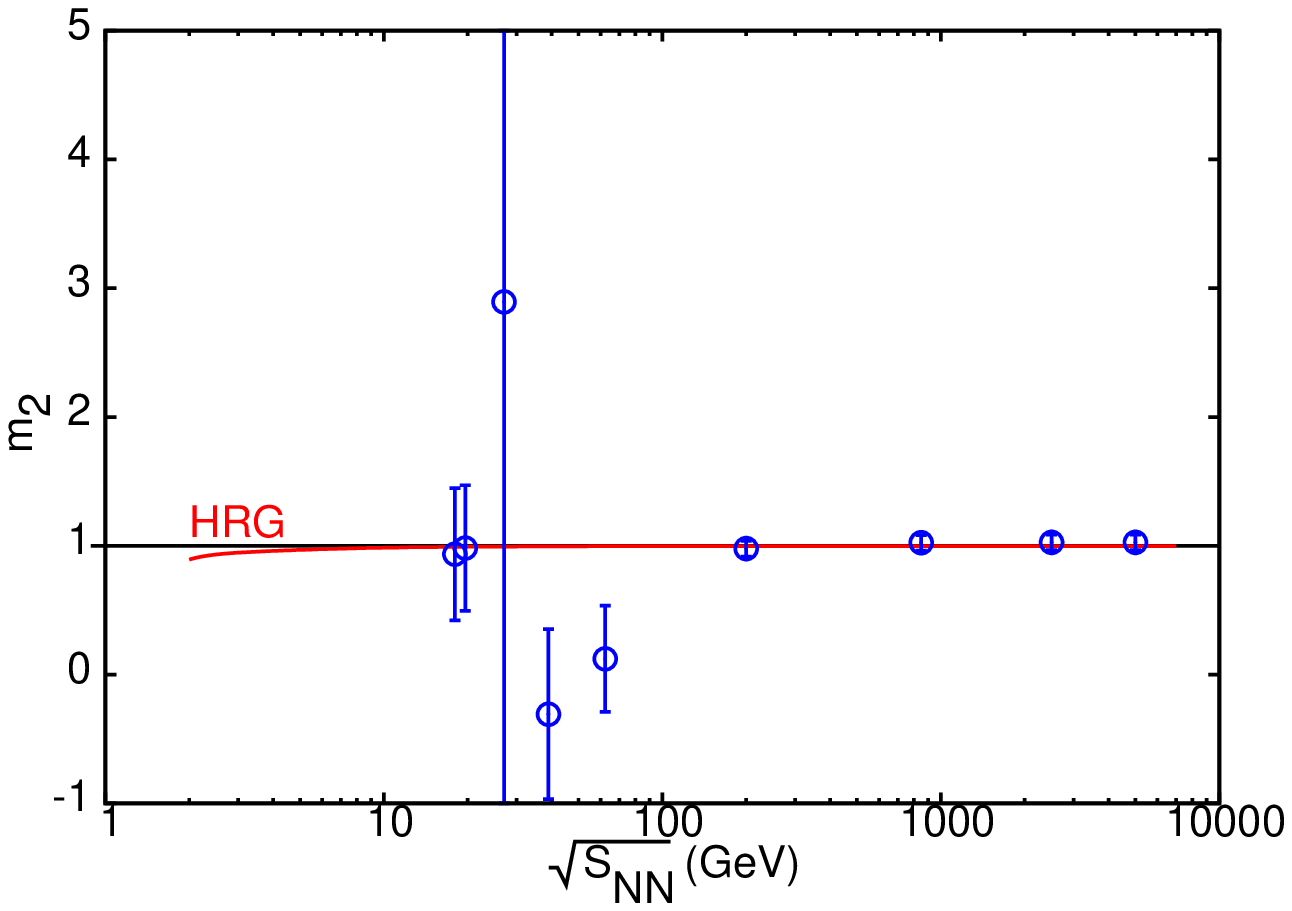}  
\includegraphics[scale=0.48]{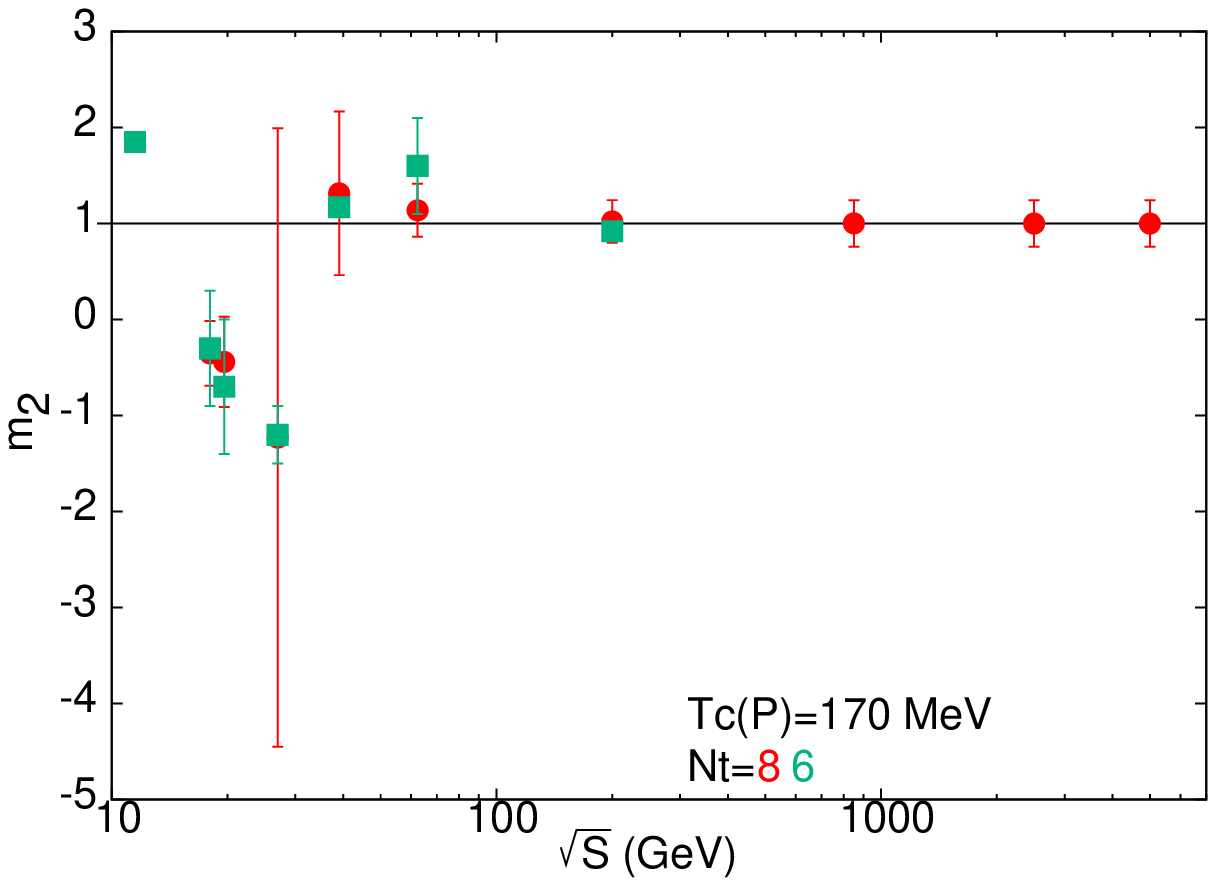} 
\caption{Results for $m_1$ (top) and $m_2$ (bottom left) on a $N_t$=6 lattice
as a function of the colliding energy. A power law fit to the high energy data
and a hadron resonance gas fit is also shown. The bottom right  panel shows
preliminary results for $m_2$ on the $N_t$=8 (green) along with the $N_t=$6
(red) lattices.  From Ref. \cite{our3, our5}.}
\label{m1m2m3}
\end{center}
\end{figure} 

The top panel and the bottom left panel in Figure \ref{m1m2m3} display the
results \cite{our3} for the ratios $m_1$ and $m_2$ on the $N_t=6$ lattice; the
results for coarser $N_t=4$ lattice are similar and can be found in Ref.
\cite{our3}.  Also shown in both these figures are results for a hadron
resonance gas which does not assume the existence of a critical point.  The
bottom right panel compares the preliminary results for $m_2$ for the $N_t=8$
\cite{our5} lattices along with those for the $N_t=6$.  A general agreement in
it is an indication again of lattice cut-off effects being rather small.  The
source of the large errors at the critical point for $N_t =8$  could very well
be the critical fluctuations themselves.  One needs to investigate this by
varying the statistics of the gauge configurations as well as the number of
vectors employed to estimate each quark propagator.   In all the figures, one
observes a smooth and monotonic behaviour for large $\sqrt{s}$ which is well
reproduced by the hadron resonance gas.  Note that even in this smooth region,
any experimental comparison is exciting since it is a direct non-perturbative
test of QCD in hot and dense environment, subject to the assumption that the
freeze-out curve $T$ and $\mu_B$ actually do correspond to those of the
fireball produced in those experiments.  Other lattice QCD predictions, such as
the equation of state or the transition temperature can be compared with
experiments only indirectly by employing them in a hydrodynamical simulation
and comparing the simulation results, e.g., for the elliptic flow, with the
corresponding experimental results.  Since initial conditions in such
hydrodynamical simulations are not uniquely known, such an indirect comparison
hinges on several other modelling aspects.

Remarkably, a non-monotonic behaviour is visible in Fig.  \ref{m1m2m3} at the
estimated critical point \cite{our2, our4} in all $m_i$, suggesting that i) an
experimental study of these ratios can possibly look for the QCD critical point
and ii) it should be accessible to the low energy scan of RHIC at BNL.  Indeed,
even if one were to be cautious in trusting the numerical precision of the
present lattice results, what should be clear is that qualitatively such a
non-monotonic behaviour is predicted by the existence of a QCD critical point;
the most relevant fact is that the $r_{ij}$'s used to compute $m_i$ had a memory
of the lattice result for the QCD critical point in the previous section.  To
that extent, any model/theory with similar memory of a critical point would lead
to a similar non-monotonic behaviour.  Moreover, if it is observed at even any
other nearby location than those estimates, it would still signal the presence
of QCD critical point, albeit somewhat shifted from the points in Fig.
\ref{CrPt}.

In order to confront these results on the baryon number fluctuations with data,
one needs to address the issue of neutral baryons---neutrons---which are not
easy to detect and are thus missed.  It turns out that proton number
fluctuations suffice \cite {HaSt}.  Since the diverging correlation length at
the critical point is linked to the $\sigma$ mode which cannot mix with any
isospin modes, and the isospin susceptibility $\chi_I$ must be regular there.
Assuming protons, neutrons, pions to dominate, Ref \cite{HaSt} showed $\chi_B$
to be dominated by proton number fluctuations only.  The STAR collaboration has
exploited this idea and constructed the ratios $m_1$ and $m_2$ from
net proton distributions \cite{STAR} in a specific rapidity  range ($|y| < 0.5$)
and a transverse momentum window (0.4 $<$ $p_{\rm T}$ $<$ 0.8 GeV/$c$).

\begin{figure}[htb]
\hskip 1.5 cm \includegraphics[scale=0.5]{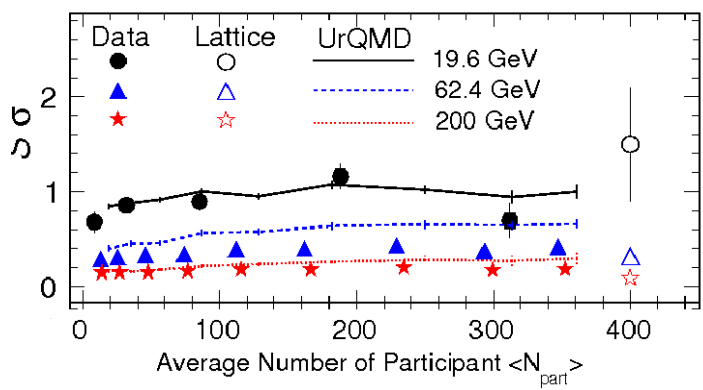}
\caption{Comparison of the STAR collaboration results with our $m_1$
results as a function of number of participants. From Ref. \cite{STAR}.}
\label{prlst}
\end{figure} 

Figure \ref{prlst} shows their results for $m_1$ ($S \sigma$ in their notation)
against the lattice data \cite{our3}. Remarkably, one observes a good agreement
with the lattice results.  Similar agreement is also seen for $m_2$. Note that
the higher colliding energy, $\sqrt{s}$, corresponds to smaller chemical
potential $\mu_B/T$.  Since higher order terms in the Taylor expansion would
clearly be much less significant for such values, the results of \cite{our3} are
even more reliable.  It is very heartening to note that this most direct test of
non-perturbative QCD with the experimental data is so good, raising the hope
that the results from the RHIC energy scan may be able to locate the critical
point this way. 

\begin{figure}[htb]
\includegraphics[scale=0.55]{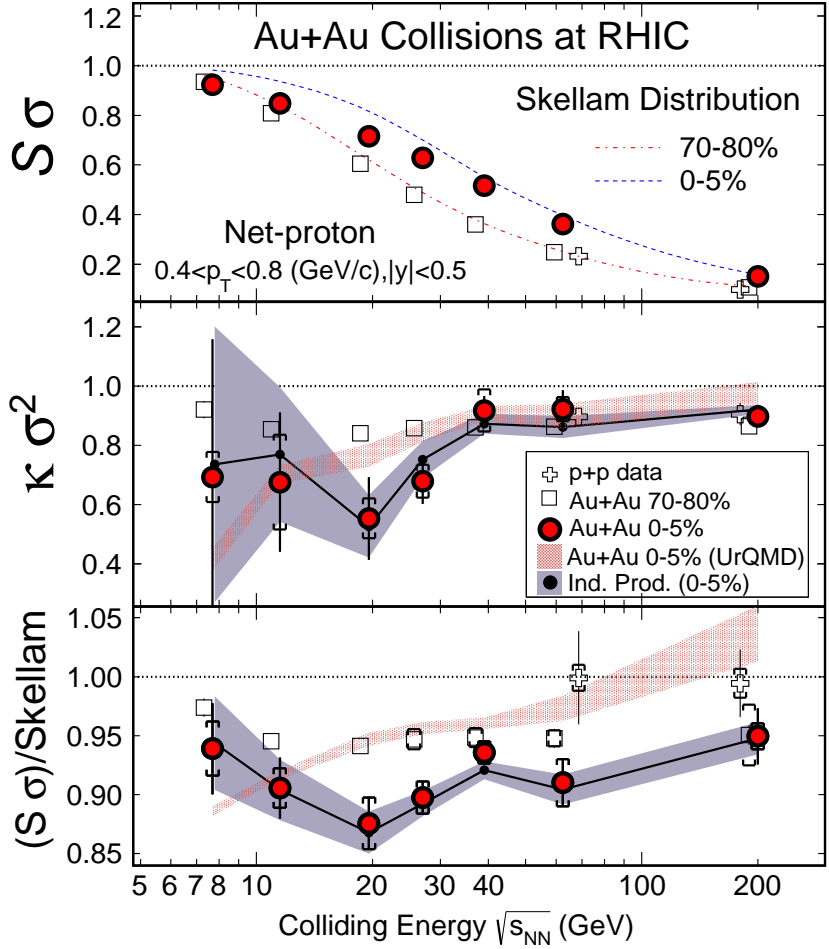} 
\caption{ The STAR results for for the equivalent of $m_1$ ($ S\sigma$) and
$m_2$($\kappa \sigma^2$) as a function of the colliding energy.  From Ref.
\cite{BeST}. For details, see the text}
\label{CrPt1}
\end{figure} 

Figure \ref{CrPt1} displays the latest results \cite{BeST} from the STAR
collaboration for the RHIC beam energy scan for $m_1$($S \sigma$ in their
notation) and $m_2$ ($\kappa \sigma^2$ in their notation). The latter should be
compared with the lattice results \cite{our3,our5} for $m_2$ in Figure
\ref{m1m2m3} (bottom right panel) for the two finest lattices. As seen in the
top panel of Fig. \ref{CrPt1}, the results for $m_1$ do increase with the
$\sqrt{s}$, as expected from the lattice results of Fig.  \ref{m1m2m3} and a
default hadron resonance gas model with no correlations.  The bottom panel
therefore exhibits the ratio of the data to the default model to check for
deviations.  The STAR results for the 0-5\% centrality for Gold-Gold collisions
(filled circles) in the two lower panels show a good qualitative and to some
extent quantitative agreement with the theoretical predictions \cite{our3,our5}
using the lattice methods.  In particular, both show a flat behaviour at large
$\sqrt{s}$ and a dip at a lower value.  For the lattice results, the dip
appears due to the QCD critical point.  Whether it is so for the experimental
data as well is a million dollar question.  One should perhaps check whether
the dip is unique at all.  The experimentalists checked this by plotting the
70-80 \% centrality data (open squares) in the same way.  These collisions are
expected to be described by simpler nucleon-nucleon like collisions and without
any extreme density deposition.  One sees no dip for these data.  Simulation
from the UrQMD model which does not incorporate any QCD critical point does not
show a dip even in the 0-5\% centrality data, as seen in Fig.  \ref{CrPt1}.  On
the other hand, it seems to do a better job for the 70-80\% centrality data.
Moreover, the analogue of $m_1$ (the top and bottom panels) also show
deviations from the expected behaviour, as more vividly seen in the bottom
panel of the figure.  While these tantalizing results are clearly very
exciting, and point towards a possible QCD critical point, more work is still
needed on the theoretical and experimental fronts in form of improved precision
as well as on dynamic model studies to pin the QCD connection more directly.

For small $z=\mu_B/T$, one may invert the relations between $m_i$'s, or similar
ratios, and {\em determine} in principle the freezeout parameters
\cite{our3,BiBnl, Bmw}  from the first principles lattice computations as well.
Alternatively, one may fit the experimental data to such ratios by varying the
scale, the cross-over temperature $T_c$, and {\em determine} it from the data
\cite {Gup11} directly.  If the lattice computations were to be really precise,
and came with a very reliable estimate of various systematic errors, any of the
above approach to extract information would be equally valid, and could even
afford a consistency check.  What I outlined above in details, is perhaps the
most conservative approach which aims to make a comparison of the lattice
predictions with the data to discern an imprint of non-perturbative QCD in the
experimental data at larger $\sqrt{s}$ and the QCD critical point at the smaller
ones.  It needed an input scale $T_c \simeq 170$-175 MeV, which appears as a
normalization in Fig. \ref{CrPt}, and was needed to relate the dimensionless
ratios appearing in the lattice computations to the freeze-out parameters
\cite{stm}.  

Recall that the lattice predictions \cite{our3,our5} in Figs. \ref{m1m2m3} and
\ref{prlst} were based on the assumption that the freeze-out curve deduced from
hadron abundances represents the true temperature and chemical potential of the
fireball when the freeze-out occurs.  Even the attempt to determine the
freeze-out parameters or the transition temperature $T_c$ from the experimental
results by using precision lattice results cannot escape from such an
assumption.  {\em A~priori}, there is no physical reason for this assumption to
be true.  Indeed, we do not even have a theoretical proof that thermalization
has to be achieved in the heavy ion collisions at all, let alone for the
existence of a freeze-out curve.  On the other hand, the data based extraction
\cite{stm}  of the freeze-out curve is indeed very impressive and provides the
best justification for it.  

\section{Summary}
\label{su}

The QCD phase diagram in the $T$-$\mu_B$ plane, Fig. \ref{CrPt}, has begun to
emerge using first principles lattice approach.  Lattice results for $N_t =6$
and 8 are first encouraging steps towards continuum limit, suggesting rather
small cut-off effects.  All lattice results so far suggest a  critical point at
a small $\mu_B/T \sim$ 1-2.  This maybe very encouraging for the beam energy
scan program at RHIC, BNL, which may be able to locate the critical point
experimentally.  Ratios of nonlinear susceptibilities appear to be smooth on
the freeze-out curve at large colliding energy. STAR results on proton number
fluctuations in Fig. \ref{prlst} appear to agree with the lattice QCD
predictions \cite{our3}, making this a unique direct non-perturbative test of
hot and dense QCD in all such experimental tests so far.  Critical point leads
to structures in the $m_i$, which may be accessible in experiments.   So far
tell-tale signs of a critical point have emerged in the RHIC experimental
results.  An interesting question is whether the deviations visible in the RHIC
energy scan in Fig. \ref{CrPt1} can uniquely be associated with the QCD
critical point, as the lattice results \cite{our3,our5} in Fig.  \ref{m1m2m3}
suggest, or whether mundane explantions can emerge for them.  Clearly, a lot of
work needs to be done by both theorists and experimentalists, apart from simply
improving the precision to resolve this.  At any rate, exciting future awaits
the beam energy scan at RHIC, and the future programs at FAIR/NICA,  as well as
all of us interested in it.

\section{Acknowledgements}
It is my great pleasure to acknowledge the wonderful role of my collaborators,
Debasish Banerjee, Saumen Datta, Sayantan Sharma, and in particular, Sourendu
Gupta from TIFR.  My work would not have been possible without the excellent
computing facility provided by ILGTI, TIFR, and maintained so well by Ajay Salve
and Kapil Ghadiali.

\end{document}